\documentclass[aps,pre,groupedaddress,twocolumn]{revtex4}
\pdfoutput=1

\newcommand{\Rc}{\mathcal{R}}
\newcommand{\Dc}{\mathcal{D}}

\usepackage{hyperref}
\usepackage{subfigure}
\usepackage{color}
\usepackage{graphicx}

\begin{document}
\expandafter\ifx\csname urlprefix\endcsname\relax\def\urlprefix{URL }\fi

\DeclareGraphicsExtensions{.pdf, .jpg}

\title{\large
   {\it N}-slit  interference: \\
   Path  integrals,  Bohmian  trajectories.  }

\large
\author{Valeriy I. Sbitnev}
\email{valery.sbitnev@gmail.com}
\affiliation{B.P.Konstantinov St.-Petersburg Nuclear Physics Institute, Russ. Ac. Sci.,
     Gatchina, Leningrad district, 188350, Russia.}


\date{\today}

\begin{abstract}
 Path integrals give a possibility to compute in details routes of particles from particle sources through slit gratings and further to detectors.
 The path integral for a particle passing through the Gaussian slit results in
 the Gaussian wavepacket.
 The wavepackets prepared on $N$ slits and superposed together give rise to interference pattern  in the near-field zone. It transforms to diffraction in the far-field zone represented by divergent principal rays, at that all rays are partitioned from each other by $(N-2)$ subsidiary rays.
 The Bohmian trajectories in the near-field zone of
 $N$-slit gratings show wavy behavior. And they become straight in the far-field zone.
 The trajectories show zigzag behavior on the interference Talbot carpet
 (ratio of particle wavelength to a distance between slits
 are much smaller than 1 and $N\gg 1$).
 Interference from the the $N$-slit gratings is simulated by scattering monochromatic neutrons (wavelength=0.5 nm).
 Also we have considered simulation of interference fringes arising at scattering on an $N$-slit grating of fullerene molecules
 (according to the real experiment stated in e-print 1001.0468). \\


\end{abstract}

\maketitle

\large

\section{\label{sec:level1}Introduction.}

 Classical mechanics operates with point particles behavior of which is found unambiguously from its variational principles~\cite{Lanczos:1970}.
 Initially, we begin with the action integral:
\begin{equation}\label{eq=1}
    S = \int\limits_{t_{0}}^{~~t_{1}} L(\overrightarrow{\bf q}, \dot{\overrightarrow{\bf q}};t)dt
\end{equation}
 Here $L(\overrightarrow{\bf q}, \dot{\overrightarrow{\bf q}};t)$ is a Lagrangian function
 equal to difference of kinetic energy and potential energy of the particle.
 Dynamical variables $\overrightarrow{\bf q}$ and $\dot{\overrightarrow{\bf q}}$
 are generalized coordinate and velocity of the particle.
 Ones proclaim, that the action $S$ remains constant along an optimal path of the movement particle.
 It is the least action principle.
 According to this principle, finding of the optimal path adds up to solution of the extremum problem ${\delta}S=0$.
 This solution leads to Hamilton-Jacobi equation (HJ-equation):
\begin{equation}\label{eq=2}
   - {{\partial S}\over{\partial\,t}} = H(\overrightarrow{\bf q}, {\overrightarrow{\bf p}};t).
\end{equation}
 Here $H(\overrightarrow{\bf q}, {\overrightarrow{\bf p}};t)=({\overrightarrow{\bf p}}\dot{\overrightarrow{\bf q}})-L(\overrightarrow{\bf q}, \dot{\overrightarrow{\bf q}};t)$ is the Hamilton function and
 ${\overrightarrow{\bf p}}={\nabla S}$ is a particle momentum.
 It is worth mentioning the optical-mechanical analogy~\cite{Lanczos:1970} of particular solutions of the HJ-equation.
 It brings to light on deep parallels between mechanical trajectories, optical rays, and even fluid streams fall under these parallels.
\begin{figure}[htb!]
  \centering
  \begin{picture}(200,170)(0,10)
      \includegraphics[scale=0.65]{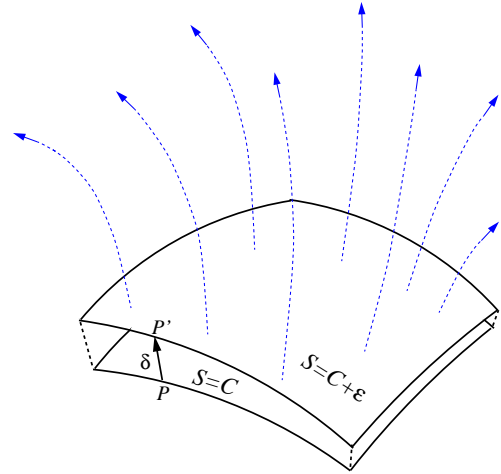}
  \end{picture}
  \caption{
  Shift of a surface $S=C$ to a new position $S=C+\varepsilon$ on a value $\delta\ll 1$~\cite{Lanczos:1970}.
  Possible trajectories shown by dotted blue curves intersect the surfaces perpendicularly.
  }
  \label{fig=1}
\end{figure}

 Please note, that the gradient of the function $S$, that is, $\nabla S$, is  directed normally to its surface $S={\rm const}$.
 Consider two nearby surfaces $S=C$ and $S=C+\varepsilon$, see Fig.~\ref{fig=1}.
 Let us trace a normal from an arbitrary point $P$ of the first surface up to its intersection with the second surface at point $P\,'$.
 Hereupon,  make another shift of the surface on the value of $2\varepsilon$, thereupon on $3\varepsilon$, and so forth.
 Until all the space will be filled with such secants.
 Normals drawn from $P$ to $P\,'$, thereupon from $P\,'$ to $P\,''$, and so forth,
 disclose possible trajectories of a mechanical system, since $|\nabla S|=\varepsilon/\delta$ represents a value of the gradient of $S$.
 This relation can be expressed in the vector form
\begin{equation}\label{eq=3}
    {\overrightarrow{\bf p}}={\nabla S}.
\end{equation}
 So far as the momentum ${\overrightarrow{\bf p}}=m{\overrightarrow{\bf v}}$ has a direction tangent to the trajectory,
 then the following statement is true~\cite{Lanczos:1970}:
 {\it trajectory of a moving point is perpendicular to the surface} $S={\rm const}$.
 Fig.~\ref{fig=1} shows possible trajectories by dotted curves intersecting the surfaces $S$ perpendicularly.

 It is appropriate to mention here the Liouville theorem, that adds to the conservation law of energy one more a conservation law.
 Meaning of the law is that a trajectory density is conserved independently of deformations of the surface that encloses these trajectories.
 Mathematically, this law is expressed in a form of the continuity equation
\begin{equation}\label{eq=4}
    {{\partial \rho}\over{\partial\,t}}
    +({\overrightarrow{\bf v}}\nabla\rho)=0.
\end{equation}
 Here $\rho$ is the density of moving mechanical points with the velocity ${\overrightarrow{\bf v}}={\overrightarrow{\bf p}}/m$.

 Thus, two equations, the HJ-equation and the continuity equation, give mathematical description of ensemble of moving points undergoing no noise.
 Draw attention here, that the continuity equation depends on solutions of the HJ-equation via the term
 ${\overrightarrow{\bf v}}={\nabla S}/m$.
 Whereas, the HJ-equation does not undergo dependence on solutions of the continuity equation.
 This is essential moment at description of moving ensemble of the classical objects in contrast to quantum ones, as we will see farther.

 In 1933 P. A. M. Dirac drew attention to a special role of the action in quantum mechanics~\cite{Dirac1933}
 that can exhibit itself in  expressions by means of a term $\exp\{{\bf i}S/\hbar\}$.
 It is appropriate to notice the following observation: the action here plays a role of a phase shift.
 And according to the principle of stationary action, the phase shift has to be constant along an optimal path of the particle.
 In 1945 Paul Dirac emphasized once again, that the classical and quantum mechanics have many general points of crossing~\cite{Dirac1945}.
 In particular, he had wrote "We can use the formal probability to set up a quantum picture rather close to the classical picture in which the coordinates $q$ of a dynamical system have definite values at any time. We take a number of times $t_{1},t_{2},t_{3},\cdots$ following closely
 one after another and set up the formal probability for the $q$'s at each of these times lying within specified small ranges,
 this being permissible since the $q$'s at any time all commutate. We then get a formal probability for the trajectory of the system
 in quantum mechanics lying within certain limits. This enables us to speak of some trajectories being improbable
 and others being likely."~\cite{Dirac1945}.

 Dirac's observations had influence to R. Feynman's searching acceptable language for description of moving quantum objects,
 where decisive role has a term $\exp\{{\bf i}L{\delta t}/\hbar\}$~\cite{Feynman1948}.
 Idea is that the above term proposes mapping a wave function from one state to another divided by a small time interval ${\delta t}$.
 Feynman's genius insight has resulted in understanding that the integral kernel (propagator) of the time-evolution operator can be expressed
 as a sum over all possible paths (not just the classical one) connecting the points $q_{a}$ and $q_{b}$
 with the weight factor $\exp\{{\bf i}S(q_{a},q_{b};t)/\hbar\}$~\cite{Grosche1993, MacKenzie2000}:
\begin{equation}\label{eq=5}
    K(q_{b},q_{a}) = \sum_{all~paths} A\exp\{{\bf i}S(q_{a},q_{b};T)/\hbar\},
\end{equation}
 where $A$ is an normalization constant.

 As a result we have now a powerful mathematical apparatus, the path integral technique~\cite{FeynmanHibbs:1965}.
 Briefly, according to this idea, there are many possible trajectories, that can be traced from a source to a detector.
 But only one trajectory, submitting to the principle of stationary action, is real.
 The others cancels each other because of interference effects.
 Such an interpretation is extremely productive at generating intuitive imagination for more perfect understanding quantum mechanics.

 The contents of the article is as follows.
 In Sect.~\ref{sec:level2} being based on the path integral technique we study migration of a particle across a single Gaussian slit.
 As a result we get a Gaussian wave packet equivalent to that used in the articles~\cite{SanzMiret2007, SanzMiret2008, Sbitnev0907}.
 Sect.~\ref{sec:level3} introduces superposition of the wave packets emitting from $N$ slits.
 Here we describe also interference from $N$-slit grating in the near-field zone and diffraction in the far-field zone.
 Sect.~\ref{sec:level4} describes the quantum HJ-equation and the continuity equation,
 that lead to the guidance equation for finding Bohmian trajectories.
 Simulation of the fullerene molecular interference stated in~\cite{JuffmannEtAl2010} is considered in Sect.~\ref{sec:level5}.
 Sect.~\ref{sec:level6} discusses virtual trajectories, virtual sources of spherical waves, and much more.
 Almost all virtual trajectories cancel each other. Remaining trajectories are Bohmian ones.
 Sect.~\ref{sec:level7}, concluding section, discusses subtle problems of virtual trajectories emerging in vacuum,
 due to which an optimal paths through $N$-slit grating are disclosed.

\section{\label{sec:level2}Gaussian slit}

\begin{figure}[htb!]
  \centering
  \begin{picture}(200,170)(30,10)
      \includegraphics[scale=0.85]{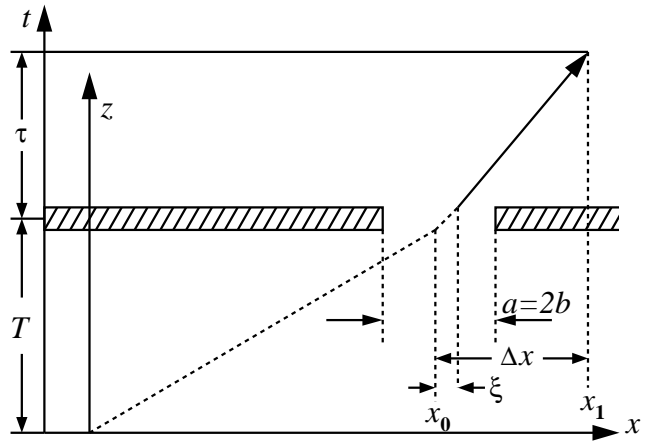}
  \end{picture}
  \caption{
  Movement of the particle through a slit~\cite{FeynmanHibbs:1965}.
  }
  \label{fig=2}
\end{figure}

 Before we will analyze interference on the $N$-slit grating, let us consider a particle passing through a single slit.
 The problem has been considered in detail in~\cite{FeynmanHibbs:1965}.
 We will study migration of the free particle along axis x, see Fig.~\ref{fig=2}.
 Its Lagrangian is as follows
\begin{equation}\label{eq=6}
    L = m\,{{\dot{x}^{2}}\over{2}}.
\end{equation}
 Here $m$ is mass of the particle and $\dot{x}$ is its velocity.
 By translating a particle's position on a small value $\delta x =(x_{b}-x_{a})\ll 1$,
 being performed for a small time $\delta t = (t_{b}-t_{a})\ll 1$,
 we find that a weight factor of such a translation has the following form
\begin{equation}\label{eq=7}
    {\rm e}^{\;{\bf i}L{\delta t}/\hbar} =
    \exp\Biggl\{ {{{\bf i}m(x_{b}-x_{a})^{2}}\over{2\hbar(t_{b}-t_{a})}}
    \Biggr\}.
\end{equation}
 Pay attention on the following situation: since argument of the exponent contains multiplication of the Lagrangian $L$ by ${\delta t}$,
 putting into operation the Lagrangian~(\ref{eq=6}) we obtain result $(x_{b}-x_{a})^{2}/(t_{b}-t_{a})$.
 Further we will see, that the weight factor~(\ref{eq=7}) plays an important role.
 Now, by means of such small increments let us trace passing the particle from a source through the slit and farther, Fig.~\ref{fig=2}.

 We suppose, that at the time $t=0$ the particle leaves the origin of coordinates $x=0$.
 Let we know,  that after a time $T$ the particle enters to the vicinity $x_{0}\pm b$ of a point $x_{0}$, see Fig.~\ref{fig=2}.
 The question is: what is the probability to disclose the particle after a time $\tau$ at a point $x_{1}$
 remote from the point $x_{0}$ at a distance $\Delta x=(x_{1}-x_{0})$?
 Let the particle outgoing from the point $x=0$ at the time $t=0$  passes a slit between the points $x_{0}-b$ and $x_{0}+b$ at the time $t=T$.
 Let us compute the probability of discovering the particle at some point $x_{1}$ after the time $\tau$, i.e., at $t=T+\tau$.
 Because of existence of an opaque screen, the problem can have no solution if we apply laws only for moving the free particles along direct paths.
 The particle should pass through the slit in order to reach the point $x_{1}$.
 In this connection, we partition the problem into two parts.
 Each part relates to movement of the free particle.
 In the first part we consider the particle moving from the point $x=0$ at the time $t=0$ to a point $x=x_{0}+\xi$,
 reaching it at time $t=T$, where $|\xi|\le b$.
 The second part deals with the particle passing through the point $x=x_{0}+\xi$ at the time $t=T$
 and moving to the point $x_{1}$, reaching it at the time $t=T+\tau$.
 A full probability amplitude is equal to integral convolution of two kernels, each describing movement of the free particle:
\begin{widetext}
\begin{equation}\label{eq=8}
    \psi(x_{1},x_{0}) = \int\limits_{-b}^{b} K(x_{1},T+\tau;x_{0}+\xi,T)K(x_{0}+\xi,T;0,0)d\xi.
\end{equation}
  Here the kernel reads
\begin{equation}
\nonumber
  K(x_{b},t_{b};x_{a},t_{a}) = \Biggl[{{2\pi{\bf i}\hbar(t_{b}-t_{a})}\over{m}}\Biggr]^{-1/2}
  \cdot \exp\Biggl\{{{{\bf i}m(x_{b}-x_{a})^{2}}\over{2\hbar(t_{b}-t_{a})}}\Biggr\}.
\label{eq=9}
\end{equation}
\end{widetext}
 It describes a transition amplitude from $x_{a}$ to $x_{b}$ for a time interval $(t_{b}-t_{a})$~\cite{FeynmanHibbs:1965}.
 Consequently, the integral~(\ref{eq=8}) computes the probability amplitude of transition from the origin $x=0$
 to the point $x_{1}$ through the all possible intermediate points $\xi$ situated within the interval $(x_{0}-b,x_{0}+b)$.

 The expression~(\ref{eq=8}) is written in accordance with a rule of summing amplitudes for successive events in time.
 The first event is the moving particle from the origin to the slit.
 The second event is the movement of the particle from the slit to the point $x_{1}$.
 The slit has a finite width. And passing through the slit is conditioned by different alternative possibilities. Therefore, we need to integrate along all the slit width in order to get a right result. All particles, moving through the slit, are free particles and their corresponding kernels are given by the expression~(\ref{eq=9}). By substituting this kernel to the integral~(\ref{eq=8}) we get the following detailed form
\begin{widetext}
\begin{equation}\label{eq=10}
    \psi(x_{1},x_{0}) = \int\limits_{-b}^{b}
    \Biggl({{2\pi{\bf i}\hbar\tau}\over{m}}\Biggr)^{-1/2}
    \exp\Biggl\{{{{\bf i}m(\Delta x-\xi)^{2}}\over{2\hbar\tau}}\Biggr\}
    \Biggl({{2\pi{\bf i}\hbar T}\over{m}}\Biggr)^{-1/2}
    \exp\Biggl\{{{{\bf i}m(x_{0}+\xi)^{2}}\over{2\hbar T}}\Biggr\}d\xi.
\end{equation}
\end{widetext}
 Integration here is executed along the slit width $a=2b$, i.e., from $-b$ to $+b$.

 Formally, ranges of the integration can be broadened from $-\infty$ to $+\infty$.
 But in this case, one should introduce step function $G(\xi)$ equal to unit in the interval $[-b,+b]$ and equal to zero outside this interval.
 Such a form-factor with sharp edges can be suitable, if wavelength of the particle is much more than interatomic distances at slit's edges.
 If a material from which the slit grating is produced shows fuzzy edges of the slits, then such a form-factor is incorrect.
 Simplest form-factor, simulating such fuzzy edges of the slits, is the Gaussian form-factor
\begin{equation}\label{eq=11}
    G(\xi) = \exp\{-\xi^{2}/2b^{2}\}.
\end{equation}
 Effective width of the curve is conditioned by a parameter $b$.
\begin{figure}[htb!]
  \centering
  \begin{picture}(200,150)(10,10)
      \includegraphics[scale=0.75]{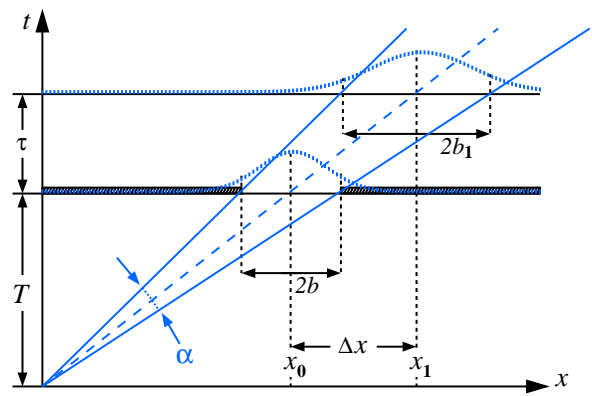}
  \end{picture}
  \caption{
  Trajectories of particles, passing through the Gaussian slit~\cite{FeynmanHibbs:1965}, form a ray with an angle $\alpha$ of the particle beam.
  }
  \label{fig=3}
\end{figure}
 For such a form-factor roughly two thirds of all its area is situated between the points $-b$ and $+b$.
 If the particles would move in classical way, then we should wait, that after time $\tau$ they will be distributed along axis $x$
 likewise as before but with a new distribution center $x_{1}$ be shifted on a value $\Delta x$ from the point $x_{0}$.
 Width $b_{1}$ of the new distribution is also broadened. The both parameters, $x_{1}$ and $b_{1}$, are determined by expressions
\begin{equation}\label{eq=12}
    x_{1} = x_{0}\Biggl(1+{{\tau}\over{T}}\Biggr), \hskip24pt
    b_{1} = b\Biggl(1+{{\tau}\over{T}}\Biggr),
\end{equation}
 as shown in Fig.~\ref{fig=3}.
 One can see, all classical trajectories, passing through the Gaussian slit, form a ray within an angle of beam $\alpha$.
 Observe, however, that quantum particles, in contrast to the classical ones, at scattering on the slit manifest wavelike nature.
 Relation between matter waves and their wavelengths had been established by de Broglie.
 Observe, that the wavelike nature underlies phase shifts of the moving particles in an observation point.

 Taking into account the above stated, Eq.~(\ref{eq=10}) can be rewritten in the following form
\begin{widetext}
\begin{equation}\label{eq=13}
    \psi(x_{1},x_{0}) = \int\limits_{-\infty}^{\infty}
    {{mG(\xi)}\over{2\pi{\bf i}\hbar\sqrt{T\tau}}}
    \Biggl(\exp
    \Biggl\{{{{\bf i}m}\over{2\hbar}}
    \Biggl[
    {{(\Delta x-\xi)^{2}}\over{\tau}} +
    {{(x_{0}-\xi)^{2}}\over{T}}
    \Biggr]
    \Biggr\}
    \Biggr)d\xi.
\end{equation}
 By substituting $G(\xi)$ from Eq.~(\ref{eq=11}) to this expression and integrating it we obtain
\begin{eqnarray}\nonumber
  \psi(x_{1},x_{0}) &&= \sqrt{{{m}\over{2\pi{\bf i}\hbar}}}
  \Biggl[T\tau\Biggl({{1}\over{T}}+{{1}\over{\tau}}+{{{\bf i}\hbar}\over{mb^{2}}}\Biggr)\Biggr]^{-1/2} \\
  &&\times \exp\Biggl\{{{{\bf i}m}\over{2\hbar}} \Biggl(
  \Biggl({{(x_{1}-x_{0})^{2}}\over{\tau}}+{{x_{0}^{2}}\over{T}}
  \Biggr) -
  {{(-(x_{1}-x_{0})/\tau+x_{0}/T)^{2}}\over
  {(1/\tau+1/T+{\bf i}\hbar/mb^{2})}} \Biggr)
   \Biggr\}.
\label{eq=14}
\end{eqnarray}
 Here $\Delta x$ has been replaced  by its explicit expression $\Delta x = (x_{1}-x_{0})$.

\subsection{\label{subsec:level2A}Solitary Gaussian slit}

 Rewrite Eq.~(\ref{eq=14}) by executing insignificant transformations
\begin{eqnarray}\nonumber
   \psi(x_{1},x_{0}) &&= \sqrt{{{m}\over{2\pi{\bf i}\hbar}}}
  \Biggl[T(1+\tau/T)\Biggl(1+{{{\bf i}\hbar\tau}\over{mb^{2}(1+\tau/T)}}\Biggr)\Biggr]^{-1/2} \\
  &&\times\exp\Biggl\{ {{{\bf i}m}\over{2\hbar}} \Biggl(
  {{(x_{1}-x_{0})^{2}}\over{\tau}}+{{x_{0}^{2}}\over{T}} -
  {(-({x_{1}-x_{0})/\tau+x_{0}/T)^{2}\tau}\over{ {(1+\tau/T)} {(1+{\bf i}\hbar\tau/mb^{2}(1+\tau/T))} }}.
  \Biggr\}
\label{eq=15}
\end{eqnarray}
\end{widetext}
 The transformation, resulting to Eq.~(\ref{eq=15}), affects only arrangement of the time parameters $\tau$ and $T$.
 Now, according to Eq.~(\ref{eq=12}) we could introduce $b_{1}=b(1+{\tau/T})$.
 But in the limit $T\rightarrow\infty$ we have equality $b_{1}=b$.
 So, we can remove everywhere in Eq~(\ref{eq=15}) the term $(1+\tau/T)$.
 Due to the limit $T\rightarrow\infty$ we have put source of the particles to infinity.
 According to this trick, we can now suppose that the particles incident to the slit grating are described by wave function that is a plane wave.
 In accordance with the limit $T\rightarrow\infty$, Eq.~(\ref{eq=15}) transforms to the following form
 \begin{eqnarray}\nonumber
\hspace{-24pt}
    \psi(x_{1},x_{0}) &&=  \sqrt{{{m}\over{2\pi{\bf i}\hbar T}}}
    \Biggl( 1+{\bf i}{{\hbar\tau}\over{mb^{2}}} \Biggr)^{-1/2} \\
    && \times
    \exp\Biggl\{ - {{(x_{1}-x_{0})^{2}/2b^{2}}\over{1+{\bf i}\hbar\tau/mb^{2}}}
    \Biggr\}.
\label{eq=16}
\end{eqnarray}

 Let us clarify correspondence of this formula with the Gaussian wave packet,
 that is described in the articles~\cite{SanzMiret2007, SanzMiret2008, Sbitnev0907}.
 This correspondence is reached by introducing an effective slit's half-width
\begin{equation}\label{eq=17}
    \sigma={{b}\over{\sqrt{2}}}={{a}\over{2\sqrt{2}}},
\end{equation}
 where double $b$, $a=2b$, is adopted as a metric width of the slit.
 Now, let us also define the complex time-dependent spreading~\cite{SanzMiret2008}
\begin{equation}\label{eq=18}
    \sigma_{\tau}= \sigma
    +{\bf i}{{\hbar\tau}\over{2m\sigma^{}}}.
\end{equation}
 This spreading consists of real and imaginary parts.
 The imaginary part, $\Delta x_{1}=\hbar\tau/2m\sigma$, emergent in Eq~(\ref{eq=14}),
 represents uncertainty of the particle position on the axis $x$.
 What is physical sense of this part?
 First, we rewrite the imaginary part in the following form $m{\Delta x_{1}}/\tau = \hbar/2\sigma$.
 Observe that $\Delta x_{1}/\tau$ is an uncertainty, $\delta v_{x}$, of a drift velocity of the particle along the axis $x$.
 Consequently, ${\delta p_{x}}=m{\delta v_{x}}$ is the uncertainty of its momentum along the same axis.
 On the other hand, let the effective slit width $2\sigma$ be uncertainty of the particle coordinate, ${\delta x}=2\sigma$ at passing across the slit.
 As a  result, we find ${\delta p_{x}}{\delta x}=\hbar$.
 One can see, it is the Heisenberg uncertainty principle~\cite{FeynmanHibbs:1965}.

 By replacing relevant expressions in Eq.~(\ref{eq=16}) by the parameters~(\ref{eq=17}) and~(\ref{eq=18})
 we get the Gaussian wave packet in the following form
\begin{eqnarray}\nonumber
    \psi(x_{1},x_{0}) &&=  \sqrt{{{m}\over{2\pi\hbar T}}}\sqrt{{{\sigma}\over{\sigma_{\tau}}}}\cdot{\rm e}^{{\bf i}3\pi/4} \\
   &&\times \exp\Biggl\{-{{(x_{1}-x_{0})^{2}}\over{4\sigma\sigma_{\tau}}}
    \Biggr\}.
\label{eq=19}
\end{eqnarray}
 Here we have rewritten the square root of imaginary unit  as a  factor $\exp\{{\bf i}3\pi/4\}=\sqrt{-{\bf i}}$.

 Pay attention to Eq.~(\ref{eq=19}), the time parameter $T$ did not disappear fully, but it remained in the denominator.
 In fact, it means, that by removing the particle source to infinity, its intensity, $\psi^{\dag}\psi$, in the vicinity
 of the slit grating become weaker in $T^{-1}$ times.
 We will suppose, that a particle beam from the source is monochromatic, i.e., velocities of all particles are equal to the same quantity $v_{z}$.
 Given path length $Z$ from the source to the slit grating, the velocity is $v_{z}=Z/T$.
 And the momentum of the particles is $p_{z}=mv_{z}$, where $m$ is its mass.

 De Broglie relations set momentum $p_{z}$ and energy $E=p_{z}^{\,2}/2m$ of the particle in correspondence with its wave characteristics
 such as a wave vector $k_{z}=p_{z}/\hbar$ and a frequency $\omega=E/\hbar$.
 In that case, a plane wave mentioned earlier can be written down as $\exp\{{\bf i}\omega t - {\bf i}k_{z}z\}$.
 Now, the Gaussian wave packet~(\ref{eq=19}) can be supplemented by this plane wave, that describes,
 in the paraxial approximation~\cite{Berry1996}, spreading of incident wave along the axis $z$:
\begin{eqnarray}\nonumber
\hspace{-24pt}
    \psi(x_{1},x_{0},z) &&=  \sqrt{{{m}\over{2\pi\hbar T}}}\sqrt{{{\sigma}\over{\sigma_{\tau}}}} \\ 
   &&\times \exp\Biggl\{-{{(x_{1}-x_{0})^{2}}\over{4\sigma\sigma_{\tau}}}
    \Biggr\} \nonumber \\
   &&\times \exp\{{\bf i}(\omega t - k_{z}z + 3\pi/4)\}.
\label{eq=20}
\end{eqnarray}
 Since in the paraxial approximation  $\tau=z/v_{z}$, then argument of the wave function contains $z$ instead of $\tau$.
\begin{figure}[htb!]
  \centering
  \begin{picture}(200,125)(20,75)
      \includegraphics[scale=0.5]{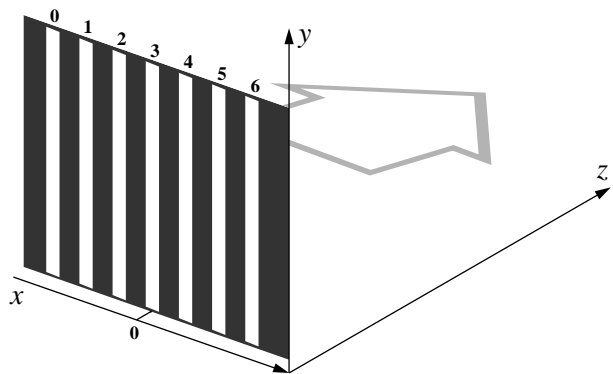}
  \end{picture}
  \caption{
 Interference experiment in cylindrical geometry.
 Slit grating with $n=0,1,\cdots,N-1$ slits is situated in a plane $(x,y)$.
 Radiation goes along axis $z$.
  }
  \label{fig=4}
\end{figure}
 In this case, interference experiment can be considered on an $N$-slit grating situated in a plane $(x,z)$, as shown in Fig.~\ref{fig=4}.

 One can see, by comparing equation~(\ref{eq=20}) with such a formula presented in~\cite{Sbitnev0907},
 that with exception of some details these formulas are equivalent.
 For example, presence of the constant phase $3\pi/4$ in Eq.~(\ref{eq=20}) is not critical for further observation of interference effects.
 Essential differences are, however, observed in normalization factors. Namely, the term
\begin{equation}\label{eq=21}
    A=\sqrt{{{m\sigma}\over{2\pi\hbar T}}}
\end{equation}
 differs essentially from analogous term in the formula presented in~\cite{Sbitnev0907}.
 Here we see the time parameter $T\gg 1$ is found in the denominator.
 As was mentioned earlier, this parameter has a deep sense:
 as you move farther the source of the particles, its intensity on the grating becomes weaker.
 For exception of this effect, the term (21) does not exert influence on the interference pattern.
 Therefore, for short we will write this normalization factor as $A$.

\section{\label{sec:level3}Ordered sequence of the slits}

 Let a screen, on which a monochromatic beam of the particle scatters,
 has $N$ slits ($n=0,1,2,\cdots,N-1$) located at equal distance from each other, as shown in Fig.~\ref{fig=4}.
 Here we have an origin of coordinates placed in the center of the slit grating.
 In this frame of reference, $n$-th slit has a position $x_{0}=(n-(N-1)/2)d$,
 where $d$ is a distance between slits.
 Further it is measured in units multiple to the wavelength $\lambda$.

 We need now to compute contributions of all paths passing from the source through all slits in the screen
 and farther to a point of observation $(x,z)$.
 Per se, we have to execute superposition in the observation point  of all Gaussian wave packets~(\ref{eq=20})
 from all slits $n=0,1,2,\cdots,N-1$.
 Such a superposition reads
\begin{eqnarray}
\nonumber
    && |\Psi(x,z) \rangle = \\
\nonumber \\
    && {{1}\over{N}}\sum\limits_{n=0}^{N-1}\psi\Biggl(x,
    \Biggl(n - {{N-1}\over{2}}\Biggr)d,z \Biggr),
\label{eq=22}
\end{eqnarray}
 and probability density in the vicinity of the observation point $(x,z)$ is
\begin{equation}\label{eq=23}
    p(x,z) = \langle \Psi(x,z) |\Psi(x,z) \rangle.
\end{equation}
 Hereinafter, the superscript $1$ at $x$ is dropped.

 Before we will take up interference effects in the near-field region and behavior of Bohmian trajectories here,
 let us consider an asymptotic limit of the formula~(\ref{eq=23}) in the far-field region.
 With this aim in mind, we preliminarily replace the term $(n-(N-1)/2)d$ in Eq.~(\ref{eq=22}) by $kd$,
 where $k$ runs $(-(N-1)/2,\cdots,(N-1)/2)$.
 Next, at summation we will neglect contribution of coefficients at $k^{2}d^{2}$ emergent at decomposition $(x-kd)^{2}=x^{2}-2kxd+(kd)^{2}$.
 The point is that the terms at $k^{2}d^{2}$ lead messed phases on infinity, due to which summation of that exponents gives zero contribution.
 Other sums with coefficients at $x^{2}$ and $2kxd$ can be easily computed.

 As a result, intensity of the particle beam in the far-field region is as follows
\begin{equation}\label{eq=24}
    I(x,z)=I_{0}(x,z)
    {{\sin\Biggl(\displaystyle {{N\zeta(x,z)}\over{2}} \Biggr)^{2}}\over
     {\sin\Biggl(\displaystyle {{\zeta(x,z)}\over{2}} \Biggr)^{2}}}.
\end{equation}
 Here terms $\zeta(x,z)$ and $I_{0}(x,z)$ read
\begin{equation}\label{eq=25}
    \left\{
      \begin{array}{cc}
        \zeta(x,z) = & \hspace{-58pt}{\displaystyle{{xd\,{\displaystyle{{z\lambda}\over{4\pi\sigma^{2}}}}}\over{2\sigma_{z}^{2}}}},\\
        \\
        I_{0}(x,z) = & {\displaystyle{{A^{2}}\over{N^{2}\,\sigma_{z}}}\exp\Biggl\{-{{x^{2}}\over{2\sigma_{z}^{2}}}\Biggr\}}.\\
      \end{array}
    \right.
\end{equation}
 The parameter $A$ is determined by Eq.~(\ref{eq=21}), and $\sigma_{z}$ has a form
\begin{equation}\label{eq=26}
    \sigma_{z} = \sigma \sqrt{1+\Biggl({{z\lambda}\over{4\pi\sigma^{2}}}\Biggr)^{2}}.
\end{equation}
 It turns, this parameter is equivalent to the instantaneous Gaussian width presented in~\cite{SanzMiret2007}.
 Emergence of the wavelength $\lambda$, in the above stated formulas,~(\ref{eq=25}) and~(\ref{eq=26}),
 is due to the de Broglie relation binding the momentum $p_{z}$
 with the wave vector $k_{z}$ of the pilot wave, namely, $\hbar k_{z}=h/\lambda=mv_{z}=mz/\tau$.

\begin{figure}[htb!]
  \centering
  \begin{picture}(200,170)(20,10)
      \includegraphics[scale=0.5]{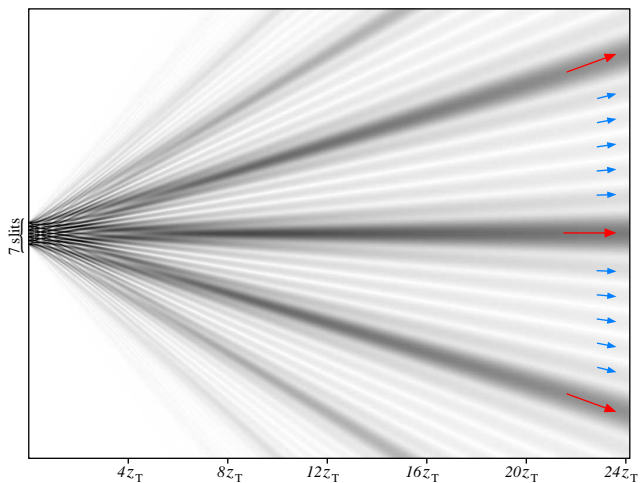}
  \end{picture}
  \caption{
 Diffraction in the far-field zone by scattering thermal neutrons ($\lambda=0.5$ nm) on $N=7$ slits grating.
 Directions of principal and subsidiary maxima  are pointed out by red and blue arrows, respectively.
 Distance between slits $d=10\lambda$, width of slits $a=2\lambda$, effective width $\sigma\approx 0.534 a$.  }
  \label{fig=5}
\end{figure}

 Let us now compare intensities of radiation from $N$ slits in the far-field region computed by Eq.~(\ref{eq=23})
 and according to the approximated formula~(\ref{eq=24}).
 Fig.~\ref{fig=5} shows, in gray palette, a distribution of the probability density~(\ref{eq=23})
 formed by radiation from $N=7$ slits grating in the far-field region.
 Thermal neutrons  are adopted here  as incident particles on the $N$-slit grating. The wavelength is $\lambda=0.5$ nm.
 Large red arrows in this figure point out directions of radiation of principal maxima having a big intensity.
 Short blue arrows, in turn, point out directions of subsidiary maxima having a low intensity.
 All nearest principal maxima are partitioned from each other by  $N-2=5$ subsidiary maxima.
\begin{figure}[htb!]
  \centering
  \begin{picture}(200,80)(0,10)
      \includegraphics[scale=0.75]{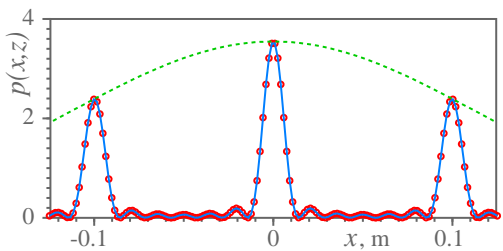}
  \end{picture}
  \caption{
 Diffraction in the far-field zone from $N=7$ slits grating, distance $z=1$ m.
 Circles relate to the probability density calculated by Eqs.~(\ref{eq=22})-(\ref{eq=23}).
 Intensity~(\ref{eq=24}) is drawn by solid curve.
 Dotted curve draws envelope $I_{0}(x,z)\cdot N^{2}$.     }
  \label{fig=6}
\end{figure}

 Fig.~\ref{fig=6} shows diffraction from $N=7$ slits grating in the far-field zone, distance $z=10^{7}z_{_{\rm T}}=1$~m.
 It is seen, that the principal maxima are partitioned from each other by $N-2=5$ subsidiary maxima.
 The curve depicted by red circles relates to the probability density, that is calculated by Eqs.~(\ref{eq=22})-(\ref{eq=23}).
 Intensity~(\ref{eq=24}) is drawn by solid curve.
 The both curves drawn in the figure give a good agreement.
 The dotted curve represents envelope $I_{0}(x,z)\cdot N^{2}$.

\section{\label{sec:level4}Madelung-Bohmian insight}

 Majority of known in physics functions are representable by linear combination of orthogonal functions,
 that are solutions of Schr{\"o}dinger wave equation~\cite{Schrodinger1926}.
 Superposition of the Gaussian wave packets from the slit grating can be represented by linear combination of orthogonal functions also.
 So, solutions of the Schr{\"o}dinger wave equation underlie this superposition as well.
 Madelung and next Bohm had shown~\cite{Madelung1926, Bohm1952a, Bohm1952b}, that the Schr{\"o}dinger equation can be reduced to two coupled equations,
 the both for real-valued functions.
 With the purpose of the following analysis of the interference effect we remind these equations.

 Almost at the same time with publication of the epochal article of Schr{\"o}dinger~\cite{Schrodinger1926},
 Madelung published his article entitled "Quantum theory in hydrodynamic form"~\cite{Madelung1926}.
 In this article Madelung found a deep analogy of behavior of the quantum system described by a wave function with the behavior of ideal liquid.
 The first step is that, the complex-valued wave function can be represented by sum of the functions, each is real-valued function.
 One of this real-valued function reduces conditionally to a rule for finding velocity of a liquid outflow.
 The other function gives conservation of density of such a quantum liquid.
 Equations, describing these two manifestations of the quantum liquid, are well known the Hamilton-Jacobi equation
 and the continuity equation
 mentioned in Introduction.
 They are Eqs.~(\ref{eq=2}) and~(\ref{eq=4}).
 A main difference exists, however.
 The difference is that the HJ-equation has an extra term, so-called the {\it quantum potential}~\cite{Bohm1952a, Bohm1952b, BohmHiley1993}.

 These renewed equations read
\begin{eqnarray}
\label{eq=27}
  -{{\partial S}\over{\partial\, t}} &=& H(\vec{\bf q},\vec{\bf p};t)+Q(\vec{\bf q},\vec{\bf p};t), \\
    {{\partial \rho}\over{\partial\,t}} &+& ({\vec{\bf v}}\,\nabla\rho)=0.
\label{eq=28}
\end{eqnarray}
 Here
\begin{equation}\label{eq=29}
    H(\vec{\bf q},\vec{\bf p};t) = {{(\nabla S)^{2}}\over{2m}} + U(\vec{\bf q})
\end{equation}
 is Hamiltonian of the quantum system, where $U(\vec{\bf q})$ is a potential energy.
 The term $Q(\vec{\bf q},\vec{\bf p};t)$ is the quantum potential
\begin{equation}\label{eq=30}
    Q(\vec{\bf q},\vec{\bf p};t) = -{{\hbar^{2}}\over{2m}}
    \Biggl[
    {{\nabla^{2}\rho}\over{2\rho}} -
    \Biggl(
    {{\nabla \rho}\over{2\rho}}
    \Biggr)^{2}
    \Biggr]
\end{equation}

 One can see, that the both equations,~(\ref{eq=27}) and~(\ref{eq=28}), are linked together.
 Namely, Eq.~(\ref{eq=27}), HJ-equation, effects to the continuity equation~(\ref{eq=28}) via the velocity $\vec{\bf v}={\nabla S}/m$.
 In turn, the continuity equation effects to the HJ-equation by means of the quantum potential,
 since the latter depends essentially  on the amplitude density $\rho$.

 Equations~(\ref{eq=27})-(\ref{eq=28}) give a rule for computing trajectories, Bohmian trajectories.
 They disclose possible paths of the particles from the source to the detector.
 The rule is given by the guidance equation that has a form
\begin{equation}\label{eq=31}
    v_{x}={\dot{x}} = {{\nabla S}\over{m}} =
    {{\hbar}\over{m}}{\Im}(|\Psi\rangle^{-1}\nabla |\Psi\rangle).
\end{equation}
 According to this equation, position $(x,z)$ of the particle in a space reaching behind the slit grating is given by
\begin{equation}\label{eq=32}
    \left\{
      {\displaystyle
      \begin{array}{cc}
        x(t) =& x_{0} + {\displaystyle\int\limits_{0}^{~~t}} v_{x}\,d\tau, \\
        \\
        z(t) =& z_{0} + v_{z}t. \hspace{24pt} \\
      \end{array}
      }
    \right.
\end{equation}

 One could think that the Bohmian trajectories are physical artifacts,
 since they enter into a rough contradiction with the Heisenberg uncertainty principle,
 because of prediction in each time moment of exact values of  coordinates and velocities of the particle.
 However, one should remember, that the Heisenberg uncertainty principle relates to procedure of measurements of complementary variables.
 It means only, that we cannot exactly measure the both variables,
 the velocity and the coordinate of the particle moving along the Bohmian trajectory.

 It should be noted, the computations~(\ref{eq=32}) are executed in the paraxial approximation, therefore the coordinate $z$ is determined by a simple shift.
\begin{figure}[htb!]
  \centering
  \begin{picture}(200,170)(27,10)
      \includegraphics[scale=0.46]{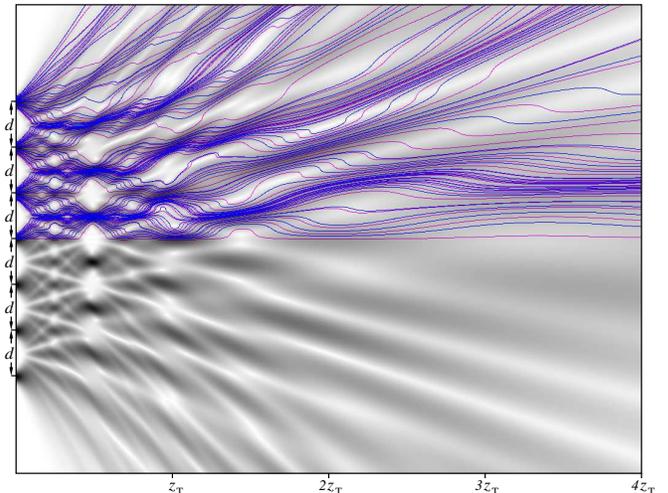}
  \end{picture}
  \caption{
  Diffraction in a zone $z\le 4z_{_{\rm T}}$ from $7$ slits grating
  of thermal neutrons ($\lambda=0.5$ nm, distance between slits $d=10\lambda$, effective slit width $\sigma\approx0.354a$).
  Wavy curves drawn by alternating violet and blue colors in the upper part
  are Bohmian trajectories divergent from the slits to infinity.
  }
  \label{fig=7}
\end{figure}
 Fig.~\ref{fig=7} shows diffraction of monochromatic thermal neutrons ($\lambda=0.5$ nm) from $N=7$ slits grating
 (distance between the slits $d=10\lambda$ and effective slit width $\sigma\approx0.354a$).
 The diffraction pattern shown in this figure covers the near-field zone, $z\le 4z_{_{\rm T}}=400$ nm,  see Fig.~\ref{fig=5}.

\begin{figure}[htb!]
  \centering
  \begin{picture}(200,140)(0,65)
      \includegraphics[scale=0.45]{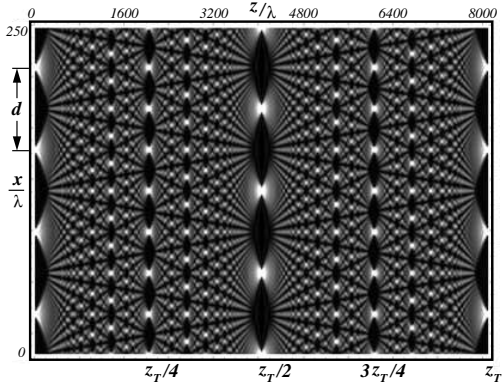}
  \end{picture}
  \caption{
  Optical Talbot effect for monochromatic light, shown as a "Talbot Carpet". The figure has been captured from
 ~{\urlprefix\url{http://en.wikipedia.org/wiki/Talbot\_effect}}.
  }
  \label{fig=8}
\end{figure}

 Here and everywhere the Talbot length
\begin{equation}\label{eq=33}
    z_{_{\rm T}} = 2{{d^{\,2}}\over{\lambda}}
\end{equation}
 is adopted as an important scale parameter that divides zones of interference patterns.
 This length starts from Henry Fox Talbot who discovered in 1836~\cite{Talbot1836} a beautiful interference pattern, Fig~\ref{fig=8},
 that carries his name.

 Fig.~\ref{fig=7} shows a probability density distribution  executed in the gray palette ranging from white to black,
 from zero density to maximal, respectively.
 It is clearly seen, that replicas of the grating begin to disintegrate already in the nearest vicinity of the slits, i.e., at $z\le 3/2z_{_{\rm T}}$.
 One can see, that at $z>2z_{_{\rm T}}$ the replicas disappear fully and transition to the far-field zone takes place.
 Namely, transformation of the probability density  distribution to a characteristic radial radiation is observed,
 see Fig.~\ref{fig=5}.

 Wavy curves shown in the upper part of Fig.~\ref{fig=7}  by alternating violet and blue colors
 are the Bohmian trajectories calculated according to formulas presented above.
 Alternation of colors permits to discern that the Bohmian trajectories do not intersect as the time $t$, i.e., $z=v_{z}t$, goes on.
 One can see, all Bohmian trajectories diverge as they move away from the slits.
 Draw attention, that the particles, passing through the slits, perform undulatory motions
 similarly to colored streams of water leaking across breakwaters.
 Such a behavior of quantum objects discloses analogy with hydrodynamic flows of liquids, as it has been described in~\cite{Madelung1926, Wyatt2005}.
\begin{figure}[htb!]
  \centering
  \begin{picture}(200,180)(27,10)
      \includegraphics[scale=0.47]{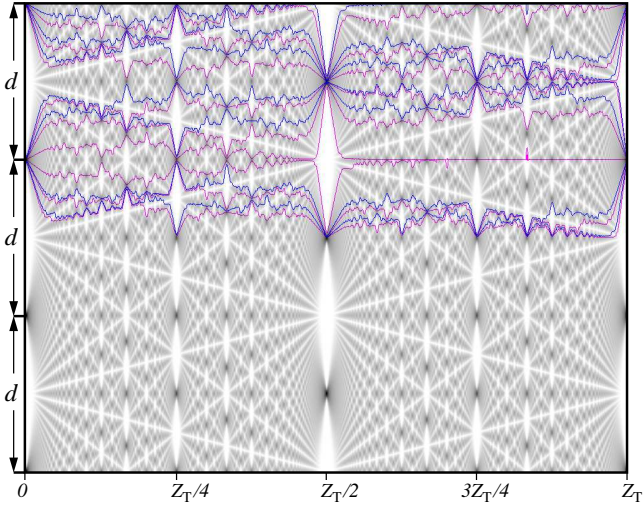}
  \end{picture}
  \caption{
  Talbot-carpet near slit sources of slit grating, $N=512$.
  Wavelength $\lambda=0.5$ nm, distance between slits $d=50\lambda$,
  effective width of slits $\sigma\approx 0.354a$.
  In the upper part Bohmian trajectories are drawn by alternating violet and blue colors.
  }
  \label{fig=9}
\end{figure}

 Since we have mentioned above Henry Fox Talbot and his beautiful pattern, Fig.~\ref{fig=8},
 it would be appropriate to consider behavior of the Bohmian trajectories on the Talbot carpet~\cite{Sbitnev0907}.
 Their behavior on the interference Talbot carpets is impressive.
 Fig.~\ref{fig=9} shows the Talbot-carpet emergent by simulation of scattering monochromatic thermal neutrons ($\lambda=0.5$ nm)
 on the slit grating containing $N=512$ slits.
 Distance between slits, in this case, is $d=50\lambda=25$ nm, so $\lambda/d=0.02\ll 1$.
 The Talbot length of that carpet is $z_{_{\rm T}}=2500$~nm. And the ratio of this length to the distance between slits is $z_{_{\rm T}}/d=100$.
 So, such Talbot carpet can be observed in a long narrow neutron guide section.

 In Fig.~\ref{fig=9} the trajectories, that we see in the upper part of the figure, demonstrate clear zigzag behavior.
 They have tendency to pass through caustics (dark patches in this figure) and avoid lacunae (white lens-like domains).
 If a trajectory  traverses across a light-colored region, it crosses this region along a shortest route.
\begin{figure}[htb!]
  \centering
  \begin{picture}(200,180)(27,10)
      \includegraphics[angle=90,scale=0.46]{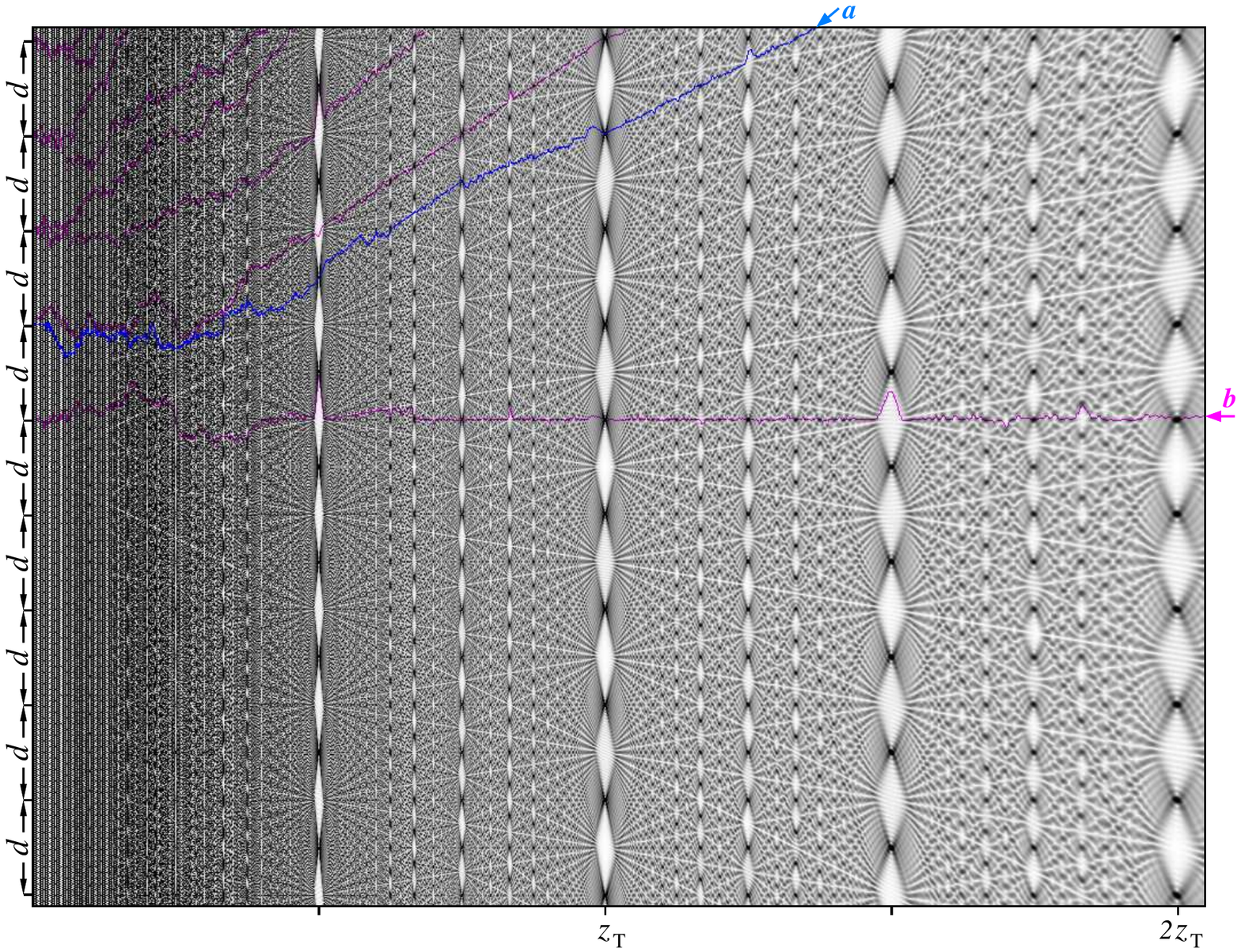}
  \end{picture}
  \caption{
  Interference pattern in the near-field region from the grating $N=64$;
  $\lambda=0.5$ nm, $d=500$ nm, $z_{_{\rm T}}=1$ mm.
  Arrows $a$ and $b$ point out to bohmian trajectories.
  }
  \label{fig=10}
\end{figure}
\begin{figure}[htb!]
  \centering
  \begin{picture}(200,180)(27,10)
      \includegraphics[scale=0.46]{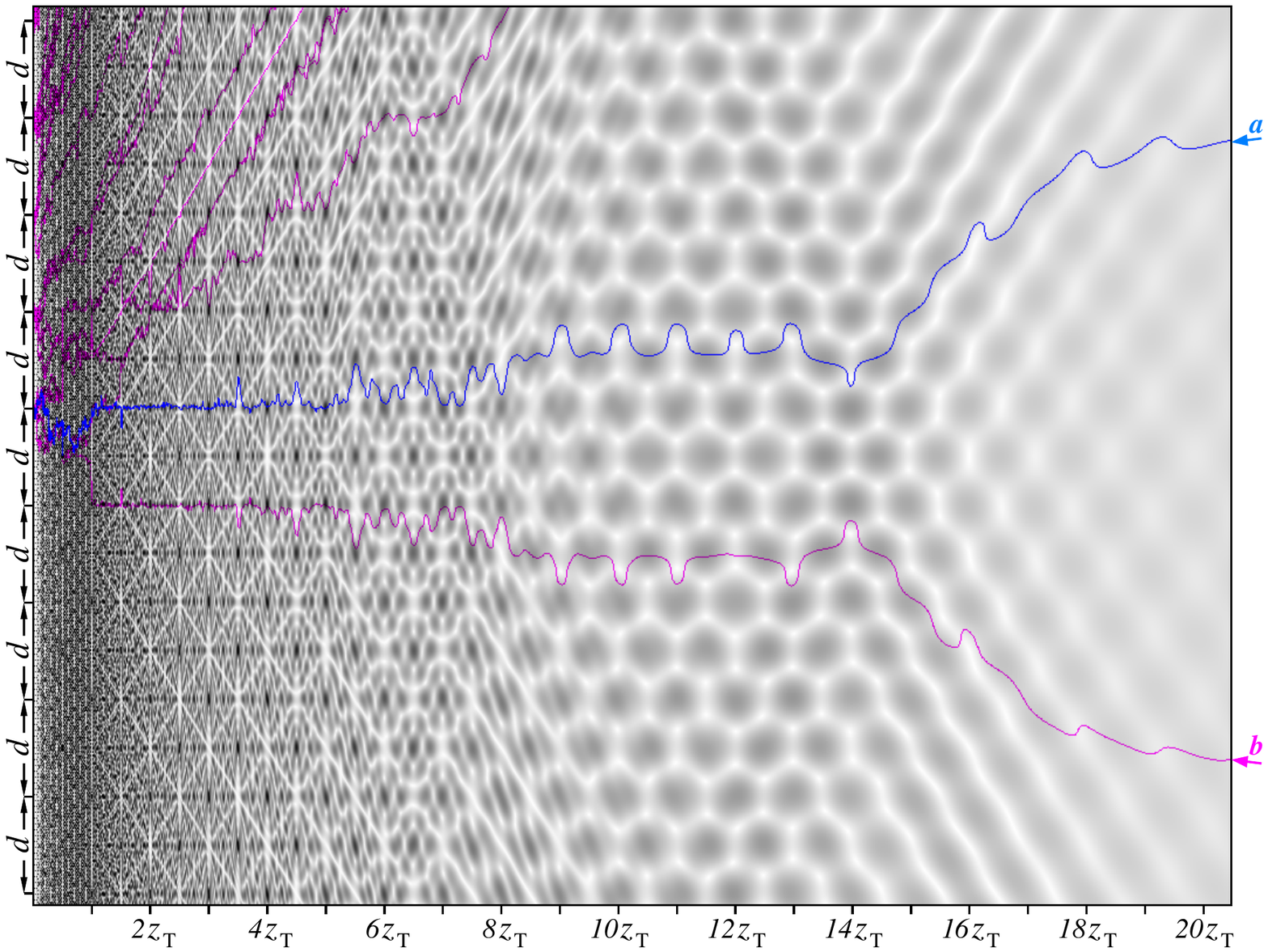}
  \end{picture}
  \caption{
  Interference pattern in the near-field region from the grating $N=64$;
  $\lambda=0.5$ nm, $d=500$ nm, $z_{_{\rm T}}=1$ mm.
  Arrows $a$ and $b$ point out to bohmian trajectories.
  }
  \label{fig=11}
\end{figure}

 In conclusion of this section let us consider a hard case of the ratio $\lambda/d$,
 namely $d=10^{3}\lambda$ and the ratio is $\lambda/d=0.001$.
 Figs.~\ref{fig=10} and~\ref{fig=11} show an interference pattern in the near-field region from the grating,
 having $N=64$ slits, and for the case $\lambda=0.5$ nm and $d=500$ nm.
 The Talbot length in this case is a very large, $z_{_{\rm T}}=10^{6}~{\rm nm}=1~{\rm mm}$.

 We see in Fig.~\ref{fig=10} the Talbot carpets, which lose their forms as  $z$ increases.
 In the vicinity of the grating the Talbot carpets demonstrate a hard fractality -
 alternation of the caustics and lacunae is so dense, that they merge into a gray mass.
 In the limit $\lambda/d\rightarrow 0$, in particular, the wave function is
 continuous everywhere, but differentiable nowhere~\cite{Sanz2005}.
 If the wave function contains fractal features, its nondifferentiability forbids a direct calculation of the trajectories.
 Tunneling processes  taking place in those cases manifest themselves as hopping events.
 Bohmian trajectories depicted in this figure show nearby the slits such a Brownian-like movement.
 As a particle moves away from the grating, its behavior becomes more predictable -
 it prefer to pass across the caustics and make the round of the lacunae, see trajectories pointed out by arrows $a$ and $b$.

 Fig.~\ref{fig=11} shows the same interference pattern, as shown in Fig.~\ref{fig=10}, but here it is extended up to $z\approx20z_{_{\rm T}}$.
 We can see, that the Talbot carpets exist only within a narrow strip by width ranging from $z=0$ to about $z=4z_{_{\rm T}}$.
 Next, they disappear, and instead of them a hexagonal packing of the interference pattern  emerges.
 It occupies a zone from about $z=9z_{_{\rm T}}$ to $z=15z_{_{\rm T}}$.
 Final stage is transition to the diffraction in the far-field zone, $z>16z_{_{\rm T}}$.
 Bohmian trajectories drawn in the upper part of the figure behave by the Brownian-like manner in the vicinity of the slit grating
 and begin to diverge from each other as $z$ increases.
 Two divergent trajectories pointed out by arrows $a$ and $b$ demonstrate transition both across the narrow strip occupied by the Talbot-carpets
 and through the hexagon zone. In the both cases we see different behavior of particles, passing through these zones.
 In the first case, sharp zigzag behavior takes place. In the second case, we observe wavy behavior.
 As the particle passes into the far-field region the wavy trajectory become more smooth, until it turns into a straight line asymptotically.

 All three Berry's conditions for emergence of the Talbot carpets shown above are satisfied.
 These conditions read~\cite{Berry1996, Berry1997, BerryKlein1996, BerryEtAl2001}:
 (a)~paraxial beam; (b)~arbitrary large number of slits; (c)~arbitrary small ratio $\lambda/d$.
 In the limits $N\rightarrow\infty$ and $\lambda/d\rightarrow 0$ the Talbot carpet tends to fractal interference carpet.

\section{\label{sec:level5}Simulation of the fullerene molecular interference}

\begin{figure}[htb!]
  \centering
  \begin{picture}(200,220)(27,10)
      \includegraphics[scale=0.74]{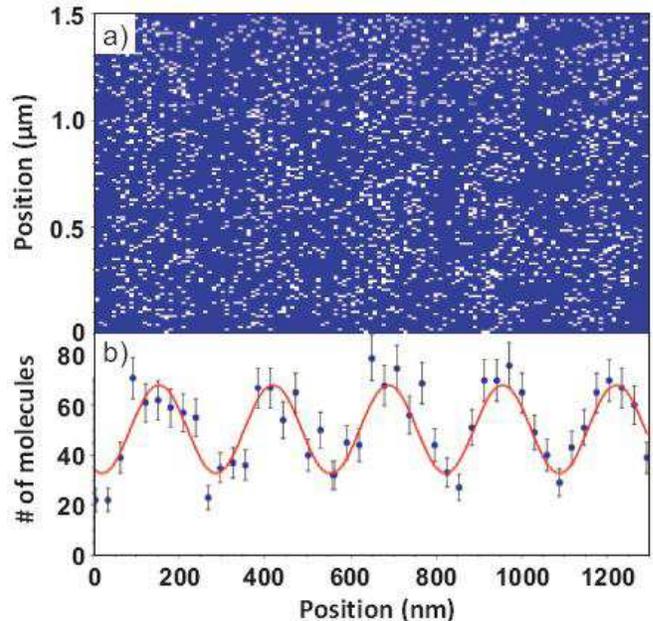}
  \end{picture}
  \caption{
  Fullerene molecular interference.
  The particle-nature (a) and the quantum wave-nature (b) of the surface deposited
  fullerene molecules in one and the same image~\cite{JuffmannEtAl2010}
  }
  \label{fig=12}
\end{figure}

 Atom and molecular interferometry~\cite{CroninEtAl2007} has an important significance in modern lithography technology.
 In this key, fullerene molecular interference experiment was recently presented in arXiv.org~\cite{JuffmannEtAl2010}.
 Heavy fullerene molecules C$_{60}$ by passing through a slit grating disclose interference fringes in the near-field zone, Fig.~\ref{fig=12}.
 White dots against blue background visible in Fig.~\ref{fig=12}(a) are images of fullerene molecules
 deposited on a silicon plate.
 All dots are seen to form vertical strips.
 Summing over each narrow  vertical band number of the dots we disclose interference fringe shown in Fig.~\ref{fig=12}(b).

 Experimental setup and parameters of the experiment are described in the article  in detail. At simulating the fullerene molecular interference fringes we will use parameters given in this article. Mass of the fullerene molecule C$_{60}$ is equal to multiplication of the nucleon mass (6 protons and 6 neutrons) to 60 nuclei. That is, the mass is $m_{_{C_{60}}}\approx 1.204\times10^{-24}$ kg. The following parameters are presented in~\cite{JuffmannEtAl2010}: (a) for C$_{60}$ with a mean velocity $v_{_{C_{60}}}=111$ m/s, a de Broglie wavelength $\lambda_{dB}$ is about 5 pm.
 Observe, that it is much smaller then radius of the fullerene, $R_{_{C_{60}}}$, that is about 370 pm!
 The SiN$_x$ grating has a highly accurate period of $d=257.40$ nm
 with open slit windows as small as 150 nm for the second grating. We take, roughly, $d=250$ nm and $a=150$ nm.

 Observe, that diffraction at each of the individual slits within the first grating expands the molecular
 coherence function, due to this it covers several slits on the second grating~\cite{JuffmannEtAl2010}.
 We assume in this key, that the fullerene molecular beam has been prepared
 as a paraxial coherent beam~\cite{CroninEtAl2007, McMorranCronin2008, JahnsLohmann1979}.
\begin{figure}[htb!]
  \centering
  \begin{picture}(200,230)(27,10)
      \includegraphics[scale=0.47]{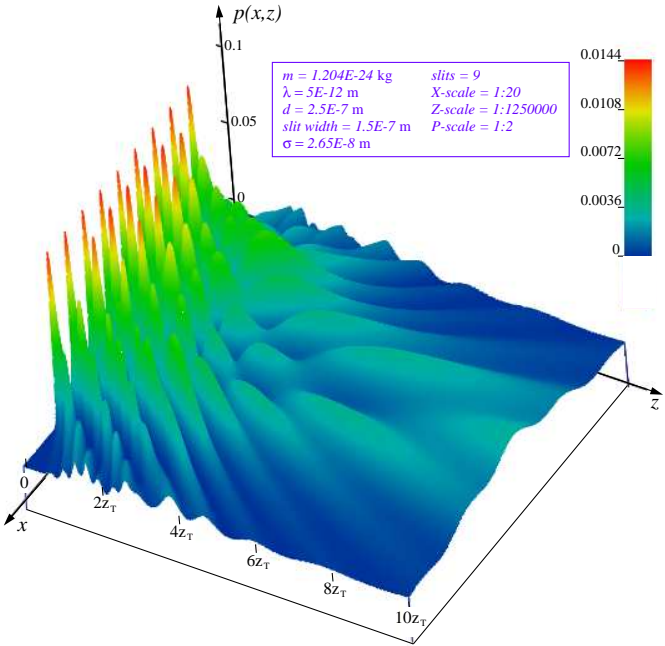}
  \end{picture}
  \caption{
  Probability density distribution of the fullerene interference simulation.
  The Talbot length is $z_{_{\rm T}}=25$ mm.
  }
  \label{fig=13}
\end{figure}
 Simulation of such a fullerene molecular beam incident on the second 9-slits grating is shown in Fig.~\ref{fig=13}.
 It is the probability density $p(x,z)$ of detecting of a fullerene molecule  within the vicinity of the point $(x,z)$.
 Parameters of the simulation are written in a table inserted in the figure.
 The Talbot length~(\ref{eq=33}) in this case is equal to $z_{_{\rm T}}=25$ mm.

\begin{figure}[htb!]
  \centering
  \begin{picture}(200,80)(20,15)
      \includegraphics[scale=0.7]{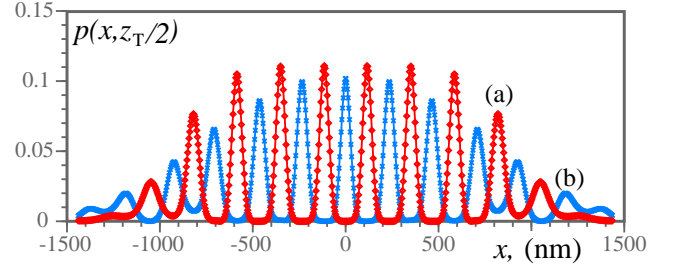}
  \end{picture}
  \caption{
  Interference fringes observed at some cross-sections of the probability density $p(x,z)$ shown in the previous figure:
  (a) $L=z_{_{\rm T}}/2=12.5$ mm; (b) $L=z_{_{\rm T}}=25$ mm.
  }
  \label{fig=14}
\end{figure}

 At half of the Talbot length, $L=z_{_{\rm T}}/2$, Juffmann {\it et al.}~\cite{JahnsLohmann1979} have observed
 the fullerene molecular interference shown in Fig.~\ref{fig=12}.
 The interference fringes, computed at the same length, are shown in Fig.~\ref{fig=14}(a).
 At comparing Figs.~\ref{fig=12}(b) and~\ref{fig=14}(a) we can see their qualitative resemblance accurate to experimental errors.
 Collisions between the fullerene molecules occur induced perhaps by dispersion of the velocities up to $\pm5$~m/s~\cite{JuffmannEtAl2010}.
 They distort the interference pattern, i.e., the pattern get blurred.
 Because of this, the experimental interference fringes have no zero level partitions.

 It would be interesting to check the next fringes positioned at the cross-section $L=z_{_{\rm T}}$.
 They should have maxima shifted on half-period, as shown in Fig.~\ref{fig=14}(b).

\begin{figure}[htb!]
  \centering
  \begin{picture}(200,180)(27,10)
      \includegraphics[scale=0.333]{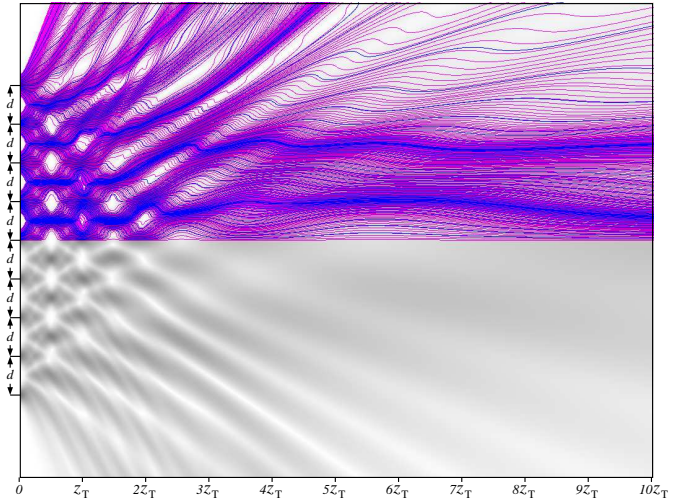}
  \end{picture}
  \caption{
  Diffraction in a zone $z\le 10z_{_{\rm T}}$ from $9$ slits grating
  of fullerene molecules beam ($\lambda=5$ pm, distance between slits $d=5\times10^{4}\lambda$).
  Wavy curves drawn by alternating violet and blue colors in the upper part
  are Bohmian trajectories divergent from the slits to infinity.
  }
  \label{fig=15}
\end{figure}
 The particle-nature of the fullerene molecules, passing through the slit grating,
 can be evaluated via the Bohmian trajectories that are calculated according to the guidance equation~(\ref{eq=31})
 and Eqs.~(\ref{eq=32}), where $v_{z}=h/\lambda m_{_{C_{60}}}\approx 110$~m/s.
 Fig.~\ref{fig=15} shows contour plot of the probability density distribution $p(x,z)$ projected on the plane $(x,z)$.
 The quantum wave-nature of the fullerene molecular flow is drawn by color gray.
 The particle-nature presented by the Bohmian trajectories is shown in the upper part of the figure.
 Predominantly, they are drawn by color violet.
 Some trajectories colored in blue are drawn against this  background.
 Due to such a color representation flows of fullerene molecules through the near-field region and further are easily visualized.
 Fullerene molecular flows are seen to have undulatory motions.

 Obviously, there are no possibilities to observe real movements of fullerene molecules through the slit grating and further~\cite{CroninEtAl2007}.
 Any perturbation of the molecule destroys its movement and, consequently, it destroys the interference pattern.
 In best case we can observe the fullerene molecular flows at crossing some kind of detecting plates
 (fixed at $z_{_{\rm T}}/2,z_{_{\rm T}},3z_{_{\rm T}}/2, \cdots$, for instance), as shown in Fig.~\ref{fig=12}(a).

\section{\label{sec:level6}Variational calculations}

 What could force the particle to carry out  such undulatory and zigzag behaviors, as shown in the figures above?
 Possible answer can be as follows:
 a set of ordered slits in the screen poses itself as a quantum object that provides a polarization of the vacuum in the near-field region.
 The polarization, in turn, induces formation of a virtual particle escort around of a flying real particle through.
 The escort "informs" the particle about the environment.
 Such an insight considering behavior of the particle with the point of view of Feynman path integrals, can be more productive,
 than fantastic ideas about splitting particles passing through the slit grating and their confluence as soon as the slits are left far behind.

 We need in this connection to continue discussion of the problems, relating to the virtual and real trajectories.
 Let us consider a number of variational procedures applied to the complex-valued function $\psi(\vec{q},\vec{p};t)$, wave function.
 It permits to see which variational scheme ends up the classical equations of movement
 and what scheme leads to the quantum mechanical equations.
 The problem, in fact, is to retrace how the virtual trajectories relate to the real trajectories, Bohmian trajectories.

\subsection{\label{subsec:level6A}Classical domain}

 Let the complex-valued function be as follows
\begin{equation}\label{eq=34}
    \psi = \sqrt{\rho}\cdot
    \exp\Biggl\{ {{{\bf i}}\over{\hbar}}
    \int\limits_{t_{0}}^{~~t_{1}} L(\vec{q}, \dot{\vec{q}};\tau)d\tau
    \Biggr\}
\end{equation}
 Let us demand that this function would retain a constant value along a path from the initial time $t=t_{0}$ to final $t=t_{1}$.
 In other words, variation of this function along this time interval has to vanish, i.e., ${\delta\psi}=0$.
 Applying this variation to the expression~(\ref{eq=34}), we disclose, that the complex-valued function ends by two equations,
 separately for real and imaginary parts.
 Each equation should vanish.

 Real part of the above equation leads immediately to the continuity equation~(\ref{eq=4}) for the probability density $\rho$.
 Imaginary part reduces to variation of the action $S$, see equation~(\ref{eq=1}),
 along paths from the initial time $t_{0}$ to the final time $t_{1}$.
 After a series of mathematical transformations~\cite{Lanczos:1970} we disclose,
 that the particle moves along an optimal path that is described by the Hamilton-Jacobi equation~(\ref{eq=2}).

 We observe in the classical case, that the particles move along the classical trajectories submitting to the principle of least action.
 By moving along the optimal paths, ensemble of the particles resembles a "cloud"  having a density distribution $\rho$.
 Evolution of the cloud obeys the continuity equation for the density distribution.

\subsection{\label{subsec:level6B}Quantum domain}

 In contrast to the previous searching of a single trajectory connecting the initial and final points,
 here all trajectories connecting these points are to be considered.
 They pass through all intermediate points belonging to a conditional set $\Rc^{3}$.
 And these all paths have to be evaluated jointly.
 Such a description goes back to the integer Chapman-Kolmogorov equation~\cite{Ventzel1975}
\begin{equation}\label{eq=35}
    p(x,z;t+s) = \int\limits_{\Rc^{n}}p(x,y;t)p(y,z;s)dy
\end{equation}
 which gives transitional probability densities of a Markov sequence.
\begin{figure}[htb!]
  \centering
  \begin{picture}(200,100)(20,15)
      \includegraphics[scale=0.8]{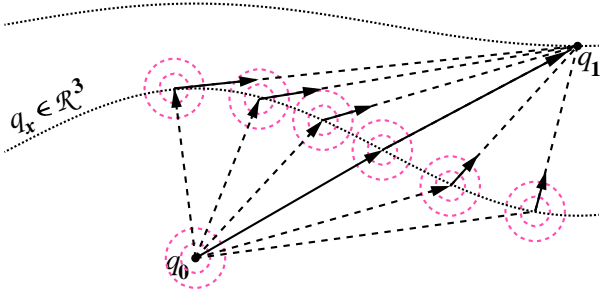}
  \end{picture}
  \caption{
  Computation of all possible paths passing from point $q_{0}$ to point $q_{1}$ through possible intermediate points $q_{x}\in\Rc^{3}$
  represents a core of the path integral method.
  }
  \label{fig=16}
\end{figure}

 Essential difference from the classical probability theory is that instead of the probabilities
 quantum mechanics deals with amplitudes containing imaginary terms, that bear phase information.
 The amplitude function charged with a phase term is a complex-valued function called the wave function.
 In this key, a transition from an initial state $\vec{q}_{0}$ to a final state $\vec{q}_{1}$ through
 all intermediate instances $\vec{q}_{\,x}$ given on a conditional set $\Rc^{3}$, see Fig. 9,
 is represented by the following path integral
\begin{eqnarray}
\nonumber
 &&\hspace{36pt} \psi(\vec{q}_{1},\vec{q}_{0},t+{\delta t}) =
\\
 &&\int\limits_{\Rc^{3}}
 K(\vec{q}_{1},\vec{q}_{x};t+{\delta t},t)\psi(\vec{q}_{x},\vec{q}_{0};t)
 \Dc^{3}{q}_{x}
\label{eq=36}
\end{eqnarray}
 in the limits ${\delta t}\rightarrow 0$ and $\vec{q}_{1}\rightarrow \vec{q}_{x}$.
 Here symbol $\Dc^{3}{q_{x}}$ represents a differential element of volume in the set $\Rc^{3}$.
 Integral kernel, i.e., propagator, $K(\vec{q}_{1},\vec{q}_{x};t+{\delta t},t)$ we suppose has a standard form
\begin{equation}\label{eq=37}
    K(\vec{q}_{1},\vec{q}_{x};t+{\delta t},t) = {{1}\over{A}}\exp\Biggl\{
    {{{\bf i}}\over{\hbar}}L(\vec{q}_{x},\dot{\vec{q}}_{x}){\delta t}
    \Biggr\}.
\end{equation}
 Standard Lagrangian in this place has a form~\cite{FeynmanHibbs:1965}
\begin{equation}\label{eq=38}
    L(\vec{q}_{x},\dot{\vec{q}}_{x}) = {{m}\over{2}}\Biggl({{\vec{q}_{1}-\vec{q}_{x}}\over{\delta t}}\Biggr)^{2}
    - U(\vec{q}_{x}).
\end{equation}
 Here $U(\vec{q}_{x})$ is a potential energy of the particle localized at the point $\vec{q}_{x}\in\Rc^{3}$.
 And $(\vec{q}_{1}-\vec{q}_{x})/{\delta t}$ is a velocity $\dot{\vec{q}}_{x}$ attached to the same point $\vec{q}_{x}$
 and oriented in the direction of the point $\vec{q}_{1}$.

 The next step is to expand terms, ingoing into the integral~(\ref{eq=36}), into Taylor series.
 For example the wave function written at the left is expanded up to the first term
\begin{equation}\label{eq=39}
    \psi(\vec{q}_{1},\vec{q}_{0},t+{\delta t}) \approx
    \psi(\vec{q}_{1},\vec{q}_{0},t) + {{\partial \psi}\over{\partial\,t}}{\delta t}.
\end{equation}
 As for the terms under the integral, here we preliminarily make some transformations.
 Let us determine a small increment
\begin{equation}\label{eq=40}
    {\vec{\xi}}=\vec{q}_{1}-\vec{q}_{x}~~\Rightarrow~~ \Dc^{3}{q}_{x}=-\Dc^{3}{\xi}.
\end{equation}
 It is believed, that a main contribution in the integral  is  only given by intermediate  sources
 from the set $\Rc^{3}$ situated  near  the point $\vec{q}_{1}$, i.e.,  corresponding to small $\vec{\xi}$.
 In this case, the Lagrangian~(\ref{eq=38}) can be rewritten as
\begin{equation}\label{eq=41}
    L(\vec{q}_{x},\dot{\vec{q}}_{x}) =
    {{m}\over{2}}{{\xi^{\,2}}\over{{\delta t}^{2}}} - U(\vec{q}_{1}-{\vec{\xi}}),
\end{equation}
 where the the potential energy $U(\vec{q}_{1}-{\vec{\xi}})$ is subjected to expansion into the Taylor series by the small parameter $\vec{\xi}$.
 The under integral wave function $\psi(\vec{q}_{x},\vec{q}_{0};t)=\psi(\vec{q}_{1}-{\vec{\xi}},\vec{q}_{0};t)$ is subjected also
 to expansion into the Taylor series up to the second terms of the expansion
\begin{eqnarray}
\nonumber
  &&  \psi(\vec{q}_{1}-{\vec{\xi}},\vec{q}_{0};t) \approx  \psi(\vec{q}_{1},\vec{q}_{0},t) \\
  && - \bigl(\nabla\psi\cdot\vec{\xi}\,\bigr) \;+\; \nabla^{2}\psi\cdot \xi^{2}/2.
\label{eq=42}
\end{eqnarray}
 Taking into account the expressions~(\ref{eq=39})-\ref{eq=42})
 and substituting theirs into Eq.~\ref{eq=36}) we get
\begin{eqnarray}
   &&\hspace{-12pt} \psi(\vec{q}_{1},\vec{q}_{0},t) + {{\partial \psi}\over{\partial\,t}}{\delta t} =
\nonumber
   -\int\limits_{\Rc^{3}}
  {{1}\over{A}}\exp\Biggl\{
    {{{\bf i}}\over{\hbar}}{{m}\over{2}}{{\xi^{2}}\over{\delta t}} \Biggr\}\times  \\
\nonumber
  &&\hspace{-12pt} \biggl(1 - {{{\bf i}}\over{\hbar}}\Bigl(U(\vec{q}_{1})
   - (\nabla U\cdot\vec\xi\,) + \Delta U\cdot\xi^{2}/2\Bigr){\delta t}\biggr)\times
   \\
  &&\hspace{-12pt} \Bigl(\psi(\vec{q}_{1},\vec{q}_{0},t)
  - \bigl(\nabla\psi\cdot\vec{\xi}\,\bigr) \;+\; \Delta\psi\cdot \xi^{2}/2
  \Bigr)\Dc^{3}{\xi}.
\label{eq=43}
\end{eqnarray}
 Here the exponent, containing the potential term $U(\vec{q}_{1}-\vec{\xi}){\delta t}$, was preliminarily expanded into the Taylor series.
 The term in this expansion having a factor of the third order of smallness, $\xi^{2}{\delta t}$, in what follows is ignored.

 The free term $\psi(\vec{q}_{1},\vec{q}_{0},t)$ represented both from the left and from the right can disappear,
 if and only if an expression, containing the constant $A$, will satisfy the following condition
\begin{equation}\label{eq=44}
    -{{1}\over{A}}\int\limits_{\Rc^{3}}\exp\Biggl\{
    {{{\bf i}}\over{\hbar}}{{m}\over{2}}{{\xi^{2}}\over{\delta t}}
    \Biggr\}\Dc^{3}{\xi} =
    -{{1}\over{A}}
    \Biggl(
    {{2\pi{\bf i}\hbar{\delta t}}\over{m}}
    \Biggr)^{3/2}
    =1,
\end{equation}
 from here it is follows
\begin{equation}\label{eq=45}
    A =
    -\Biggl(
    {{2\pi{\bf i}\hbar{\delta t}}\over{m}}
    \Biggr)^{3/2}
\end{equation}
 The power 3 emerges here because that the integration is fulfilled on the 3-dimensional set $\Rc^{3}$.
 In addition to this observation, we need to clarify integration of the terms
 $\bigl(\nabla\psi\cdot\vec{\xi}\,\bigr)$ and $\Delta\psi\cdot \xi^{2}$.
 With this aim in the mind, we mention the following two integrals~\cite{FeynmanHibbs:1965}
\begin{equation}\label{eq=46}
    {{1}\over{A}}\int\limits_{\Rc^{3}}\exp\Biggl\{
    {{{\bf i}}\over{\hbar}}{{m}\over{2}}{{\xi^{2}}\over{\delta t}}
    \Biggr\}\vec\xi \Dc^{3}{\xi} = 0.
\end{equation}
 and
\begin{equation}\label{eq=47}
    -{{1}\over{A}}\int\limits_{\Rc^{3}}\exp\Biggl\{
    {{{\bf i}}\over{\hbar}}{{m}\over{2}}{{\xi^{2}}\over{\delta t}}
    \Biggr\}\xi^{2} \Dc^{3}{\xi} = {{{\bf i}\hbar}\over{m}}{\delta t}.
\end{equation}
 In accordance with the first integral, contributions of the terms ${\nabla\phi}$ and ${\nabla U}$ in the expression~(\ref{eq=43}) disappear.
 Whereas, the second term, ${\Delta\psi}$, gains the factor $({\bf i}\hbar{\delta t}/m)/2$.

 Taking into account the above stated, we return to Eq.~(\ref{eq=43}) and write out only terms,
 that have the factors of the first order of smallness regarding ${\delta t}$
\begin{eqnarray}
\nonumber
  &&\hspace{-12pt} \psi(\vec{q}_{1},\vec{q}_{0},t) + {{\partial \psi}\over{\partial\,t}}{\delta t} \;=\;  \psi(\vec{q}_{1},\vec{q}_{0},t) \\
  && - {{{\bf i}{\delta t}}\over{\hbar}}U(\vec{q}_{1})\psi(\vec{q}_{1},\vec{q}_{0},t)
  + {{{\bf i}\hbar{\delta t}}\over{2m}}\Delta\psi.
\label{eq=48}
\end{eqnarray}
 So far, we retained here the free term $\psi(\vec{q}_{1},\vec{q}_{0},t)$.
 Meanwhile, other terms, having orders of smallness ${\delta t}^{2}$ and smaller, are excluded from this equation.
 Finally, by reducing  $\psi(\vec{q}_{1},\vec{q}_{0},t)$ from the left and from the right, we come to the Schr{\"o}dinger equation
\begin{equation}\label{eq=49}
  {\bf i}\hbar  {{\partial \psi}\over{\partial\,t}} =
  -{{\hbar^{2}}\over{2m}}\Delta\psi + U(\vec{q}_{}\,)\psi,
\end{equation}
 describing the wave field $\psi$ in the configuration set $\Rc^{3}$.
 This set evolves by force of infinitesimal transformations ($\xi\rightarrow 0$)
 by shifting in the time (${\delta t}\rightarrow 0$), see Fig.~\ref{fig=16}.
 Obviously, the subscript 1 now can be dropped.

 It should be noted, since at derivation of the Schr{\"o}dinger equation we use a condition of smallness $|\vec{q}_{1}-\vec{q}_{x}|\ll 1$,
 then decay of radiation by secondary sources with distance is not take into account.
 That is, in this derivation only a small set of the secondary sources, nearest to the point ${\vec q}_{1}$,
 give real contribution to the path integral~(\ref{eq=36}).
 The distant sources give so small contribution, that we neglect them.

 Now, let us substitute into Eq.~(\ref{eq=49}) the wave function $\psi$
 to be represented by multiplication of
 the amplitude part ${\rho}^{\,1/2}$ on the phase exponent $\exp\{{\bf i}S/\hbar\}$, see Eq.~(\ref{eq=34}).
 Next, by separating the imaginary and real parts of the Schr{\"o}dinger equation~\cite{Sbitnev:2009b}
 we come to equations~(\ref{eq=27})-(\ref{eq=28}). They are the quantum HJ-equation and the continuity equation, respectively.

\section{\label{sec:level7}Conclusion: virtual and real trajectories}

 Let us return to Fig.~\ref{fig=16}.
 Spherical waves outlined in this Figure demonstrate radiation from each point ${\vec q}_{x}$ through which virtual trajectories can pass.
 Their wavelengths are congruous with that of the particle.
 Those spherical waves radiated from each point ${\vec q}_{x}\in\Rc_{3}$, reaching ${\vec q}_{1}$, create in this point an interference effect.
 This effect shows whether a real particle passes through the point ${\vec q}_{1}$.
 If this event can happen, though, what kind is chance of it?

 One can see, that we outlined here the Huygens-Fresnel principle~\cite{Longhurst1970}.
 It proclaims that each point ${\vec q}_{x}$ at an advanced wave front is in fact the center of a fresh disturbance
 and the source of a new wave radiation.
 And the advancing wave as a whole may be regarded as the sum of all the secondary waves arising from points in the medium already traversed.
 All the secondary waves are coherent, since they are activated from the one source given in ${\vec q}_{0}$.

 It is important to emphasize, that all rays from such secondary sources represent in fact virtual trajectories
 passing from the sources up to the point ${\vec q}_{1}$.
 Along with the other virtual trajectories generated by the other secondary sources,
 all together they create in the point ${\vec q}_{1}$ an averaged effect of contribution of these secondary sources.
 This averaged effect shows whether a real particle passes by this route and what probability of this event can be.

 One can imagine, that all physical space is filled by such secondary sources of spherical waves.
 These sources are virtual in that sense, that they manifest themselves via zero-point oscillations of vacuum.
 It means, that there are permanent creations and annihilations of virtual pairs "particle-antiparticle".
 Lifetime of such pairs are associated with their total energy by the Heisenberg uncertainty principle,
 i.e., the larger uncertainty of the total energy  the smaller the lifetime of the pair.
 The same can be said about their scattering from each other:
 if the position is known perfectly, then the momentum is completely unknown, and vice versa.

 Imagine now, let an area of scattering of such pairs be equal to about an effective width of the slit, $\sigma$, that is multiple of the wavelength. In that case, the momentum, ${p}=h/\sigma$, can be sufficient to force the particle to change a direction of scattering.
 Virtual spherical waves in the vicinity of the slit can effect to behavior of the particle passing through the slit.
 Ensemble of such virtual waves, each attributed to its slit, creates a polarized effect similar to the domino effect.
 In other words, in a space, adjoining to the slit grating, zero-point oscillations of vacuum are quite  ordered.
 The slits in effect control the vacuum fluctuations.
 It is not analogous to ordered positions of atoms in crystals.
 The ordering of virtual oscillations of vacuum, in the case of the slit grating,
 is expressed via organizing the wave function that describes transport of the particles through the grating.
 The wave function represents itself the de Broglie pilot wave~\cite{Valentini0811, StruyveValentiny2009}
 that guides a particle from a source to a detector.
 A path from the source to the detector, called the Bohmian trajectory,
 is resulted from solution of the guidance equation~\cite{Wyatt2005}.

\begin{acknowledgments}

 The author thanks V.~Lozovskiy and A.~I.~Ioffe for a series of useful remarks.
 The author expresses his sincere thanks to Miss Pipa, the site administrator of  Russian Quantum Portal, for developing and writing a program calculating the density distribution of the wave function at scattering particles on $N$-slit grating. The program draws also Bohmian trajectories outgoing from the slits.

\end{acknowledgments}





\end{document}